Forecasting the Performance of US Stock Market Indices During COVID-19: RF vs LSTM

Reza Nematirad, Amin Ahmadisharaf, and Ali Lashgari[†]

**ABSTRACT**

The US stock market experienced instability following the recession (2007-2009). COVID-19 poses a significant challenge to US stock traders and investors. Traders and investors should keep up with the stock market. This is to mitigate risks and improve profits by using forecasting models that account for the effects of the pandemic. With consideration of the COVID-19 pandemic after the recession, two machine learning models, including Random Forest and LSTM are used to forecast two major US stock market indices. Data on historical prices after the big recession is used for developing machine learning models and forecasting index returns. To evaluate the model performance during training, cross-validation is used. Additionally, hyperparameter optimizing, regularization, such as dropouts and weight decays, and preprocessing improve the performances of Machine Learning techniques. Using high-accuracy machine learning techniques, traders and investors can forecast stock market behavior, stay ahead of their competition, and improve profitability.

______________________________________________________________________[†]

Kansas State University, Manhattan, KS, USA.

1. **INTRODUCTION**

The financial market is a dynamic and constantly evolving field, with stock investment playing a crucial role in wealth creation. In the domain of cognitive computer engineering, the prediction of stock prices based on historical data and textual inputs has become an important area of research. The economic instability that followed the recession of 2007-2009 in the US stock market has made investors and traders more cautious in their investment decisions. Additionally, the COVID-19 pandemic has added a significant challenge to US stock traders and investors.

As a result, there has been a surge in the use of artificial intelligence algorithms for scanning vast amounts of real-time economic and stock data, which is being used for investment decision-making. The application of Machine learning (ML) in financial market prediction has proven to be a transformative technology, successfully applied in forecasting future trends in finance and economics. This paper highlights the use of machine learning techniques in predicting the US indices of stock market, S&P 500 and Russell 2000 (RUT). It aims to help traders and investors mitigate risks and improve profits by developing machine learning models that account for the effects of the pandemic.

The study uses two machine learning models, including Random Forest and long-short term memory (LSTM). The performance of the models is evaluated using various metrics to determine the effectiveness of the approach in predicting the behavior of the S&P 500 and Russell 2000 indices.

## 1.2. RELATED WORKS

Kim et al. (2020) proposed a model for predicting the economic impact of epidemic trends using LSTM algorithms, which are known for their high performance in time series forecasting. The proposed model was validated using historical infectious disease data and achieved a 77% accuracy in predicting inflation rates. The study predicted the COVID-19 trend and its future economic impact for the next year using the proposed model. However, it noted that predicting certain economic indicators such as the stock market index may not have been appropriate with too many variables involved.

Khattak et al. (2020) aimed to explore the potential predictors of the European financial market during the COVID-19 crisis. They analyzed 21 different potential internal and external determinants using a machine learning approach called LASSO regression. The study found that before the pandemic announcement by WHO, Europe was hit by a range of predictors, including gold prices, the EUR/USD exchange rate, Dow Jones index, Switzerland, Spain, France, Italy, Germany, and Turkey indices. However, after the pandemic announcement, only France and Germany indices were selected by the LASSO approach as the most important predictors of the European market. They found that the European market was mostly affected by indices belonging to Singapore, Switzerland, Spain, France, Germany, and the S&P500 index. The LASSO approach was found to be beneficial over traditional regression methods as it allowed for variable selection and regularization in the analysis while catering to the issue of limited data during the crisis. However, the hyperparameters tuning and regularization are not addressed.

Ayala et al. (2021) proposed a hybrid approach to generate trading signals by applying a technical indicator combined with a machine learning approach. linear model, artificial neural network, random forests, and support vector regression were tested to select the most suitable machine learning technique. The triple exponential moving average and moving average convergence/divergence were considered as technical strategies for trading. The proposed hybrid trading strategy showed improved trading signals and competitiveness of the proposed trading rules. Among all the fitted machine learning techniques, linear model and artificial neural network performed best. The hybridization of technical analysis rules with machine learning models not only improved the profit but also decreased the number of trades as well as the risk of losses. However, model optimization is not addressed in that study.

Masih et al. (2022) analyzes the impact of the COVID-19 pandemic on the stock market of major IT companies such as Google, Microsoft, Apple, and Amazon. The research has used machine learning algorithms such as support vector machine and LSTM on stock market data to predict the stock prices of these companies. The study has also used the autoregressive moving average forecasting method to predict the stocks of the four companies. The results have shown that some businesses of these companies have declined during the pandemic, leading to a fall in stocks. However, some other segments such as online shopping, cloud computing, and streaming video have come out to be winning corporate strategies to fight the negative economic effect of COVID-19 and to stabilize the situation of stocks in the coming months. However, hyperparameters tuning and regularization are not taken into account.

Based on the previous studies, it is clear that machine learning techniques have been used to analyze the impact of COVID-19 on the stock market. In this paper, we focus on the use of random forest and LSTM models to predict the behavior of the S&P 500 and Russell 2000 indices after the big recession. The major contributions of this study are as follows:

- Incorporating the impact of the Covid-19 by different sample size.
- Hyperparameters tuning of the machine learning models.

This allows us to obtain a more accurate representation of the impact of the pandemic on the stock market.

## 2. METHODOLOGY

### 2.1. Data preprocessing

In order to achieve more accurate and reliable results when analyzing data, preprocessing is required to ensure the data is clean, complete, and in the proper form for machine learning (Kotsiantis, Kanellopoulos, & Pintelas, 2006). Based on the dataset, the preprocessing may include working with missing data or outliers, scaling and normalization, feature engineering, data decomposition, temporal aggregation, and etc. In this study, data cleaning, and normalization are done based on the nature of the collected data. In the real life, the collected data are most likely to have incorrect, corrupted, incorrectly formatted, duplicate, or incomplete data. Data preprocessing is an important procedure to detect and remove these data points to improve the data's accuracy, consistency, and reliability.

Data cleaning can be summarized as the following steps:

- Detecting the incorrect data points.
- Refining the incorrect data points.
- Transforming the data set to keep the order.
- Validating the refined data that meets the requirements of the data analysis.

**2.1.1 Data normalization**

Data normalization is a process in order to transform data into common scale without changing the difference in the range of the values. Data normalization is useful for transforming data into a proper form for machine learning or other analytical methods. In the literature, there are several data normalization techniques, However, this study uses Mean normalization that is one of the well-known scaling data methods. Mean normalization is an approach to scale and center values, so that the mean and variance are zero and one respectively (Alpaydin, 2020). Mean normalization transform the actual data points into a new dataset where each data points are between minus one and positive one with zero mean.

**2.2. Machine learning algorithms**

**2.2.1. LSTM**

LSTM (Hochreiter & Schmidhuber, 1997) is a type of recurrent neural network that is capable of modeling complex temporal patterns in sequential data, such as stock prices. Unlike regular feed-forward neural networks, which only consider the current input, information in the RNN travels in loops from layer to layer, preserving the context based on previous inputs and outputs (Elman, J. L. 1990). However, RNNs have some limitations such as slow computation time and difficulty retaining information over long periods (Bengio, Y., Simard, P., & Frasconi, P. 1994). LSTM overcomes these shortcomings by using a cell to remember information over time intervals and three gates to regulate the flow of information into and out of the cell. The capacity to capture long-term dependencies, versatility in handling different forecasting jobs, and handling missing variables are all advantages of LSTM in forecasting. Disadvantages include complexity in training and optimization, difficulty in interpreting results, sensitivity to hyperparameters, and potential for overfitting (Goodfellow, I., Bengio, Y., & Courville, A., 2016). The specific advantages and disadvantages may vary depending on the use case and dataset.

**2.2.2. Random Forest**

Random forest is a technique in which decision trees are used together to do classification and regression. Bagging (bootstrap aggregation) reduces variance and sustains lower bias in the Random Forest (Pentreath, 2015). Due to its ability to average different tree projection outcomes, a random forest can soothe overfitting when using training data, which makes it more accurate than a single decision tree. Random forest is advantageous for forecasting due to its high accuracy in handling complex and non-linear datasets, robustness to outliers and missing values, feature importance insights, and scalability for large datasets (Pentreath, 2015).

The random forest algorithm includes the following steps:

**Step 1:** A random sample is taken from the provided dataset by the algorithm.

**Step 2:** With each sample selected, the algorithm will create a decision tree, and subsequently, it will calculate the prediction result based on the decision trees.

**Step 3:** Every anticipated outcome will then be polled. For a classification problem, a mode is used, and for a regression problem, a mean is used.

**Step 4:** Finally, the algorithm selects the most polled prediction as a final prediction

The use of multiple algorithms can assist in identifying the underlying patterns in the data. By comparing the results of different algorithms, it is possible to get insight into how they perform on the same data and achieve better accuracy.

**2.3. Hyperparameters tuning**

Hyperparameters are the parameters of a given machine learning that cannot be tuned by the learning process, instead they must be adjusted before training. The accuracy of the machine learning algorithms highly depends on the hyperparameters and tuning them is a critical step.

**2.3.1. Randomized search**

Among the hyperparameters tuning techniques, randomized search has shown an excellent performance. From a specific distribution, the hyperparameters of a given algorithm are tuned randomly (Bergstra & Bengio, 2012).

All steps of the randomized search can be summarized as follows,

- Specifying the hyperparameters and their corresponding distribution and ranges.
- Setting the number of iterations
- Evaluate the accuracy of the model by the selected hyperparameters.
- Repeat the previous steps until the best model is found.

**2.3.2. Bayesian optimization**

LSTM can perform better when its hyperparameters are optimized using the Bayesian optimization technique. It selects the most promising collection of hyperparameters to examine using a probabilistic model and iteratively changes its model as new data come in (Bergstra, Bardenet, Bengio, & Kégl, 2011). Bayesian optimization can be used to adjust LSTM hyperparameters including batch size, learning rate, learning rate per layer, and number of LSTM layers, among others. Bayesian optimization can help boost the LSTM model's accuracy and generalization performance in a specific forecasting task by identifying the best hyperparameters.

**2.3.4. Cross-Validation**

When analyzing time series data, where observations are arranged chronologically and may be connected with one another, common cross-validation procedures assume that data points are independently and identically distributed (Hyndman & Athanasopoulos, 2018). Utilizing a rolling window is one popular method for the time series cross-validation, where the training set is made up of all data up to a specific time point and the test set is made up of the following data window. By using rolling window cross-validation, it enables the model to consider how the underlying patterns of the data change over time, which is crucial for time series forecasting. The cross-validation on a rolling basis starts from a small subset (see Fig. 1) or window of the train data and forecasts the next data point(s). The forecasted data point(s) is transferred to the next window as a new train data to forecast the new data point.

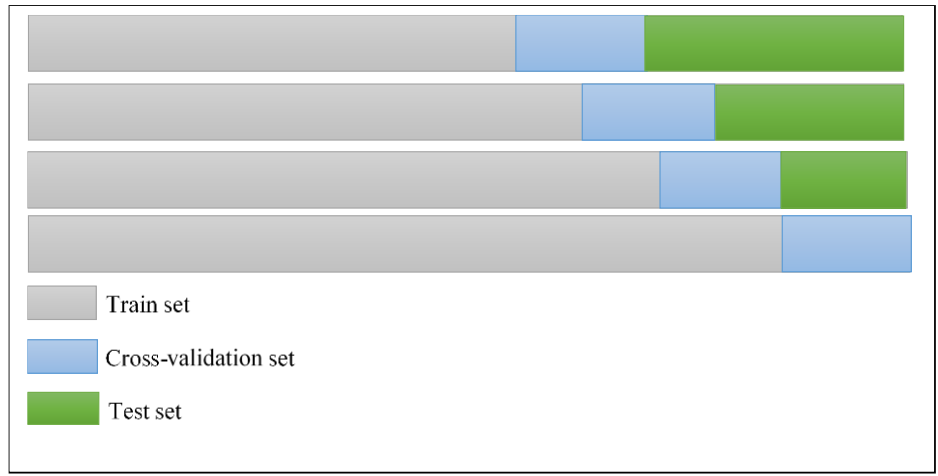

Fig. 1. Rolling basis time series data cross validation.

### 2.4. Evaluation Metrics

Evaluation metrics are used to measure the performance of a machine learning model. In the context of predicting the behavior of the stock market during the COVID-19 pandemic, two evaluation metrics are used in this study. Mean squared error (MSE) measures the average squared difference between the predicted and actual values. A lower MSE indicates better prediction accuracy (Alpaydin, 2020). R-squared ($R^2$) is a statistical measure that represents the proportion of the variance in the dependent variable that is explained by the independent variables. A higher $R^2$ indicates a better fit of the model (James, Witten, Hastie, & Tibshirani, 2013). It should be noted that using multiple evaluation metrics can provide a completer and more accurate picture of machine learning model performance. Different evaluation metrics show different aspects of model performance and can reveal model strengths and weaknesses that may not be evident with only one criterion.

## 5. RESULTS AND DISCUSSION

### 5.1. Data collection and preprocessing

The S&P 500 and Russell 2000 are both stock market indices used to measure the performance of the US equity market. The S&P 500 is generally considered a benchmark for the overall performance of the US stock market, while the Russell 2000 is often used as an indicator of small-cap stocks, which are thought to be riskier but potentially offer higher returns. The dataset used in our study includes historical data of the S&P 500 and Russell 2000 stock market indices, from June 1st, 2009 to March 31st, 2023. This time period covers a range of economic conditions, including the aftermath of the global financial crisis and the ongoing COVID-19 pandemic. In order to evaluate the impact of the COVID-19 pandemic on the S&P 500 and Russell 2000 forecasting, three different observations based on spikes of COVID-19 are created. The first one covers a stable economy before COVID-19, the second includes the pandemic's early stages, and the third extends the analysis to a longer time period impacted by the pandemic. For simplicity, the first, second, and third datasets are named D1, D2, D3 and are illustrated on Fig. 2.

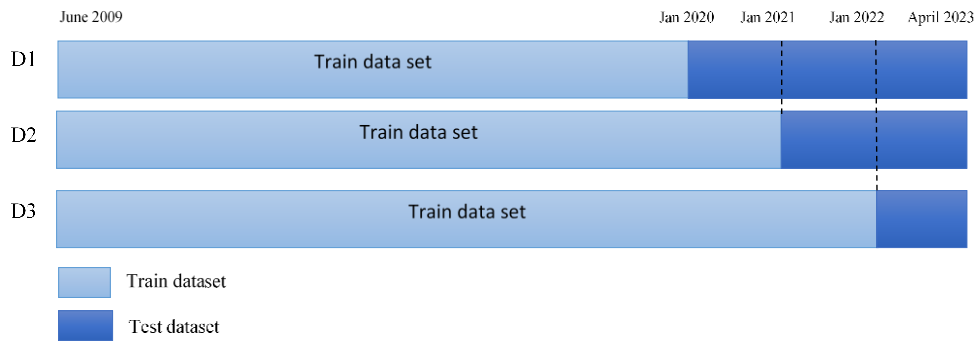

Fig. 2. Different train sample size based on the Covid-19 time period.

Further, data cleaning including detecting the NANs and date format correction are done to prepare the data for the application of machine learning. After data cleaning, the mean normalization technique is utilized to scale the data.

**5. 2. machine learning**

After cleaning the data and transforming them to an acceptable format, the next step is to feed them into machine learning to detect hidden patterns between the data points. However, the hyperparameters of the LSTM and random forest must be adjusted in advance. By taking advantage of the randomized search and Bayesian optimization, the optimal hyperparameters of the LSTM and random forest are provided in Table I and II, respectively for different sample size.

Table I. Optimal hyperparameters of the developed random forest model.

| index | Samples | Window size | Max_depth | Max-features | Min_samples_leaf | Min_samples_split | N_estimators | $R^2$ | MSE |
|---|---|---|---|---|---|---|---|---|---|
| S&P | D1 | 2 | 20 | Log2 | 1 | 2 | 217 | 0.66 | 755.2 |
| | D2 | 2 | 14 | Log2 | 3 | 2 | 213 | 0.74 | 467.5 |
| | D3 | 1 | 10 | None | 4 | 4 | 196 | 0.93 | 63.13 |
| RUT | D1 | 2 | 20 | Auto | 1 | 13 | 363 | 0.75 | 317.89 |
| | D2 | 2 | 20 | None | 3 | 6 | 356 | 0.81 | 91.93 |
| | D3 | 1 | 30 | Log2 | 3 | 2 | 165 | 0.89 | 36.84 |

Table II. Optimal hyperparameters of the developed LSTM model.

| index | Samples | Window size | Epoch | L1 | L2 | Batch size | Number of units | $R^2$ | MSE |
|---|---|---|---|---|---|---|---|---|---|
| S&P | D1 | 5 | 1 | 0.1 | 0.04 | 32 | 32 | 0.94 | 58.37 |
| | D2 | 5 | 2 | 0.1 | 0.05 | 32 | 32 | 0.96 | 55.56 |
| | D3 | 6 | 4 | 0.1 | 0.05 | 32 | 32 | 0.98 | 53.60 |
| RUT | D1 | 1 | 1 | 0.1 | 0.01 | 32 | 32 | 0.89 | 54.82 |
| | D2 | 2 | 2 | 0.1 | 0.01 | 32 | 32 | 0.97 | 42.00 |
| | D3 | 3 | 3 | 0.1 | 0.05 | 32 | 32 | 0.98 | 33.15 |

**5.2.1. Random forest**

The $R^2$ and MSE for forecasting with random forest are provided in Table I. According to the random forest model, optimal hyperparameters differ by the sample size and time period for the S&P 500 dataset. For D1, which represents the time period before the COVID-19 pandemic, the $R^2$ and MSE are 0.66 and 755.2, respectively. As of D2, the first year following COVID-19 the $R^2$ and MSE are 0.74 and 767.5, respectively. This indicates an improvement in the model performance as the S&P 500 data one year after Covid-19 is added to the training dataset. Overall, the model's performance seems to have improved as the training dataset was moved away from the initial COVID-19 dataset. Besides, Fig. 3, provides the results of forecasted versus actual values of the S&P 500 with consideration for dataset D3.

The results obtained from the random forest model for the RUT dataset are presented in table I as for different sample sizes (D1, D2, and D3). $R^2$ for D1 is 0.75, indicating that the model can explain 75% of the variance in the data. The MSE value of 317.89 is relatively high compared to other datasets (D2 and D3), indicating that the predicted values deviate significantly from the actual values. In D2, the model performance has improved in comparison with D1, with R2 of 0.81 and a lower MSE of 91.93. The best performance is achieved by D3, where the R2 value is 0.89 and the MSE is 36.84. Besides, Fig. 4, provides the results of forecasted versus actual values of the Rut with consideration for dataset D3.

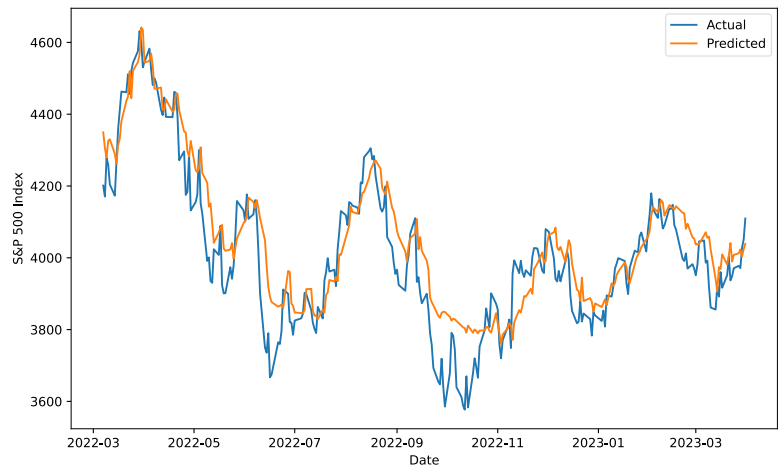

Fig. 3. Forecasted and actual values of S&P 500 with random forest for dataset D3.

**5.2.2. LSTM**

Table II shows the $R^2$ and MSE for forecasting with LSTM. Accordingly, For the three datasets (D1, D2, and D3), the $R^2$ values are all relatively high, with the lowest value being 0.94 (D1) and the highest value being 0.98 (D3). This shows that the LSTM model fits the data well and can account for variation. Besides, The MSE values for all three datasets are relatively low, with the lowest value being 53.60 and the highest being 58.37. This demonstrates that the LSTM model is able to predict the S&P 500 accurately and that it has a low average prediction error. Accordingly, the LSTM model performs well on the S&P500 dataset as the Covid-19 period time data are involved in the train datasets. Moreover, the predicted and actual values of S&P 500 and Rut, considering dataset D3, is illustrated in Fig. 5 and 6, respectively.

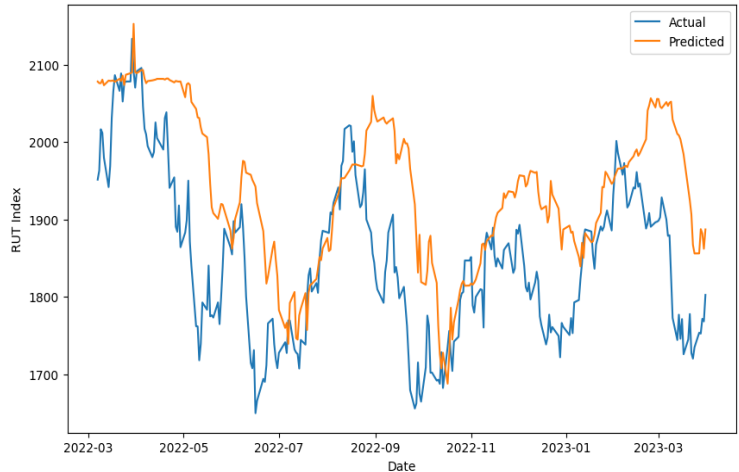

Fig. 4. Forecasted and actual values of Rut data with random forest for dataset D3.

The LSTM model on the RUT dataset shows an improvement in performance compression to random forest results. Besides, the performance of LSTM increases as the Covid-19 time period data adds to the train sample size. For example, the $R^2$ and MSE improve 9 and 0.39 percent respectively.

To sum up, as a result of the COVID-19 pandemic, the stock market experienced instability. LSTMs and random forests can perform better when the COVID-19 time period is included in the dataset. For instance, in the S&P 500 dataset, we can see that the LSTM and random forest model performance increased from D1 to D3. This can be attributed to the fact that the COVID-19 pandemic had a major effect on the stock market. Therefore, incorporating this time period in the training dataset allowed the models to capture the particular patterns and trends during this time. similarly, we can see that as we move from D1 to D3 of the RUT dataset, the performance of the LSTM and random forest models improves. As a result, disregarding the COVID-19 period when developing machine learning models can result in forecasts that are distorted. This is because the algorithm might not be able to account for changes that the pandemic-related stock market has brought about.

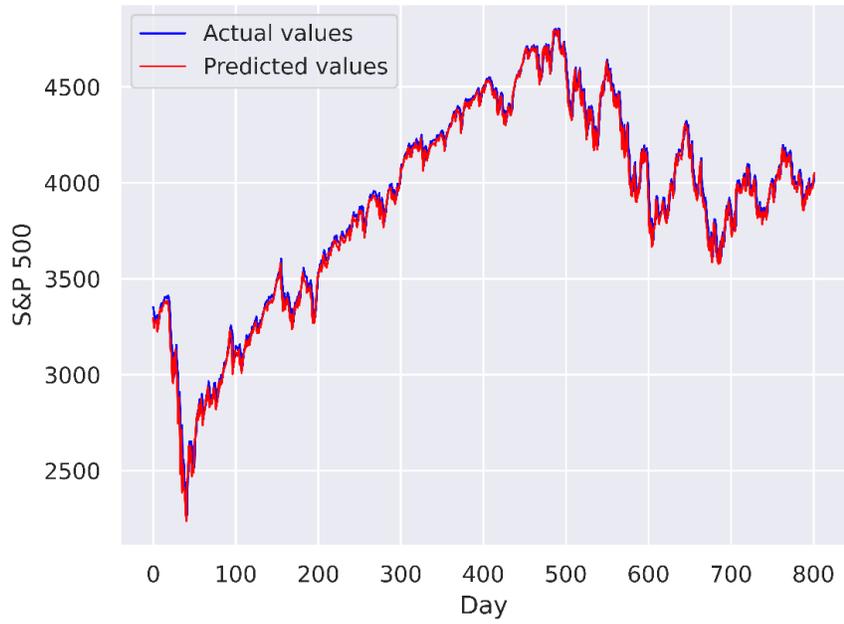

Fig. 5. Forecasted and actual values of S&P 500 with LSTM for dataset D3.

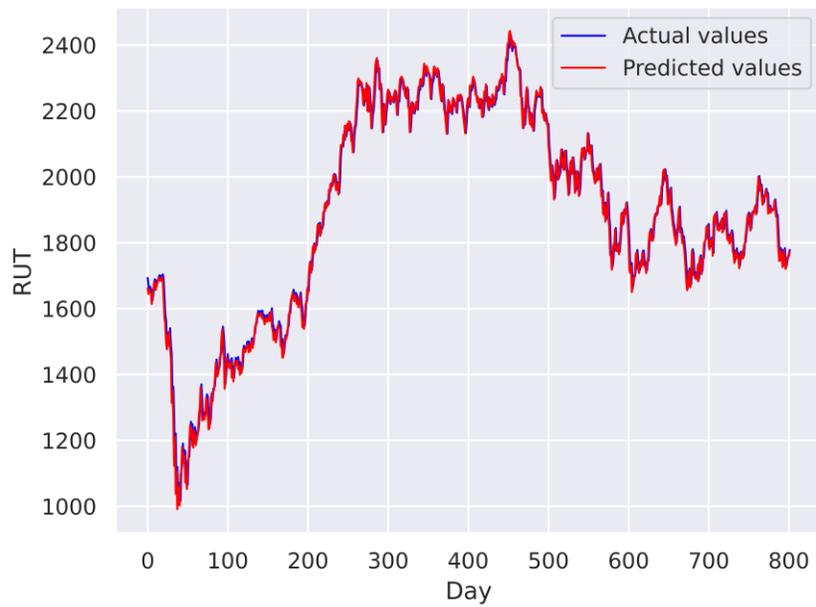

Fig. 6. Forecasted and actual values of Russell 2000 with LSTM for dataset D3.

## 6. CONCLUSION

The US stock market has been experiencing instability since the recession of 2007-2009, and the COVID-19 pandemic has posed a significant challenge to US stock traders and investors. To mitigate risks and improve profitability, traders and investors need to keep up with the stock market using forecasting models that account for the pandemic effects. This study presents the use of two machine learning models, Random Forest and LSTM, to forecast two major US stock market indices, the S&P 500 and Russell 2000. Model performance is evaluated using various metrics, including R2 and MSE. Hyperparameter optimization, regularization, and preprocessing techniques were applied to improve model performance. Simulated results show that learning accuracy increases as more Covid-19 time periods are incorporated into the process. Furthermore, based on the results, LSTM outperforms random forest in all cases. This study shows that high-accuracy machine learning techniques can be used to forecast stock market behavior and trends. As a part of future studies, the different economical indices can be combined to have a comprehensive analysis.